\documentclass[prl,twocolumn,superscriptaddress,showpacs]{revtex4}

\usepackage{graphicx}
\usepackage{amsmath}

\begin{document}

\newcommand{\SZFKI}{Research Institute for Solid State Physics and Optics,
                    POB 49, H-1525 Budapest, Hungary}

\newcommand{\GPS}{Groupe de Physique des Solides, CNRS UMR 75-88,
                  Universit\'es Paris VI at VII, Tour 23,
                  2 place Jussieu, 75251 Paris Cedex 05, France}

\title{Nucleation and Bulk Crystallization in Binary Phase Field Theory}

\author{L\'aszl\'o Gr\'an\'asy}
  \affiliation{\SZFKI}

\author{Tam\'as B\"orzs\"onyi}
  \affiliation{\SZFKI}
  \affiliation{\GPS}

\author{Tam\'as Pusztai}
  \affiliation{\SZFKI}

\date{\today}

\begin{abstract}
  We present a phase field theory for binary crystal nucleation. In
  the one-component limit, quantitative agreement is achieved with
  computer simulations (Lennard-Jones system) and experiments
  (ice-water system) using model parameters evaluated from the free
  energy and thickness of the interface. The critical undercoolings
  predicted for Cu--Ni alloys accord with the measurements, and
  indicate homogeneous nucleation. The Kolmogorov exponents deduced
  for dendritic solidification and for ``soft-impingement'' of
  particles via diffusion fields are consistent with experiment.
\end{abstract}


\maketitle

Understanding alloy solidification is of vast practical and
theoretical importance. While the directional geometry in which the
solidification front propagates from a cool surface towards the
interior of a hot melt is understood fairly well, less is known of
equiaxial solidification that takes place in the interior of the
melt. The latter plays a central role in processes such as alloy
casting, hibernation of biological tissues, hail formation, and
crystallization of proteins and glasses. The least understood stage of
these processes is nucleation, during which seeds of the crystalline
phase appear via thermal fluctuations. Since the physical interface
thickness is comparable to the typical size of critical fluctuations
that are able to grow to macroscopic sizes, these fluctuations are
nearly all interface. Accordingly, the diffuse interface models lead
to a considerably more accurate description of nucleation than those
based on a sharp interface \cite{diff,diff1}.

The phase field theory, a recent diffuse interface approach, emerged
as a powerful tool for describing complex solidification patterns such
as dendritic, eutectic, and peritectic growth morphologies
\cite{phf}. It is of interest to extend this model to nucleation and
post-nucleation growth including diffusion controlled
``soft-impingement'' of growing crystalline particles, expected to be
responsible for the unusual transformation kinetics recently seen
during the formation of nanocrystalline materials \cite{nano}.

In this Letter, we develop a phase field theory for crystal nucleation
and growth, and apply it to current problems of unary and binary
equiaxial solidification.

Our starting point is the free energy functional
\begin{eqnarray}
F=\int d {\bf r} \left\{ {\frac{\epsilon^2 T}{2}} (\nabla \phi )^2 + 
f (\phi, c) \right\},
\end{eqnarray}
\noindent developed along the lines described in \cite{phf1,phf2}. Here 
$\phi$ and $c$ are the 
phase and concentration fields, $f(\phi,c) = WTg(\phi) + [1-P(\phi)] f_S + 
P(\phi) f_L$ is the local free energy density, $W = (1-c)W_A + cW_B$ 
the free energy scale, the quartic function $g(\phi) = \phi^2 (1-\phi)^2 /4$ 
that emerges from density functional theory \cite{dft} 
ensures the double-well form 
of $f$, while the function $P(\phi)  = \phi^3 (10 - 15\phi + 6\phi^2)$ 
switches on and off the solid and liquid contributions $f_{S,L}$, 
taken from the ideal solution model. (A and B refer to the constituents.)

For binary alloys the model contains three parameters $\epsilon$, 
$W_A$ and $W_B$ that reduce to two 
($\epsilon$ and $W$) in the one-component limit. 
They can be fixed if the respective interface free energy $\gamma$, 
melting point $T_{f}$, and interface thickness $\delta$ are known 
\cite{param}. Such information is available for the
Lennard-Jones, ice-water, and Cu--Ni systems \cite{param1}, 
offering a quantitative test of our approach. 

Relying on the isothermal approximation the time evolution 
is described by Langevin equations $\partial \phi/\partial t = 
- M_{\phi} (\delta F/\delta \phi) + \zeta_{\phi}$ and $\partial c/\partial t = 
\nabla [M_c \nabla (\delta F / \delta c)] + \zeta_{\nabla c}$, where 
$(\delta F/\delta x)$ stands for the functional derivatives 
($x = \phi, c$), $M_x$ are mobilities, while $\zeta_{\phi}$ and 
$\zeta_{\nabla c}$ are appropriate noises added to RHS to mimic thermal 
fluctuations. The dimensionless form of these equations \cite{diml} 
is obtained by measuring length and time in units 
$\xi$ and $\xi^2/D_L$, $t = \tilde t \xi^2/D_L$, ${\bf r} = 
\tilde {\bf r}/\xi$. Here $\xi$ and $D_L$ are
the characteristic length scale and the diffusion coefficient 
in the liquid, while the quantities with tilde are dimensionless.

The critical fluctuation (nucleus) is a non-trivial time-independent 
solution of the governing equations. For spherical symmetry 
(a reasonable assumption), the phase field equation reduces 
to $\nabla^2 \phi = \Delta \mu (\phi, c)/(\epsilon^2 T)$. 
Here $\Delta \mu (\phi,c) = $ $WTg'(\phi) + 
[(1-c) \Delta f_A + c \Delta f_B] 
P'(\phi)$ is the local chemical potential difference relative 
to the initial liquid, prime stands for differentiation 
with respect to the argument, the local concentration is related 
to the phase-field as $c(\phi) = c_{\infty} e^{-y}/(1-c_{\infty} + 
c_{\infty} e^{-y})$, where $y =  v (W_B - W_A) g(\phi)/R + 
v(\Delta f_B - \Delta f_A) [P(\phi) - 1]/RT$, while $\Delta f_i$ 
are the volumetric free energy differences between the pure 
liquid and solid phases. Solving these equations numerically 
under boundary conditions that prescribe bulk liquid properties 
far from the fluctuations ($\phi \rightarrow 1$, and $c 
\rightarrow c_{\infty}$ for $r \rightarrow \infty$), and 
zero field-gradients at the center, 
one obtains the free energy of critical fluctuation as $W^* = 
F - F_0$. Here $F$ is obtained by numerically evaluating Eq. (1) 
after having the time-independent solutions inserted, while $F_0$ 
is the free energy of the initial liquid. This is compared with 
$W^* = (16\pi/3) \gamma_f^3/\Delta f^2$ from the sharp 
interface ``droplet'' model of the classical nucleation theory 
\cite{k91}, where $\gamma_f = \gamma(T_f)$. 

The homogeneous nucleation rate is calculated as $J = J_0 exp
\left\{-W^*/kT \right\}$, where the nucleation prefactor $J_0$ of the
classical kinetic approach is used \cite{gunton}, which proved
consistent with experiments \cite{k91}.

\begin{figure}
\resizebox{60mm}{!}{
\includegraphics{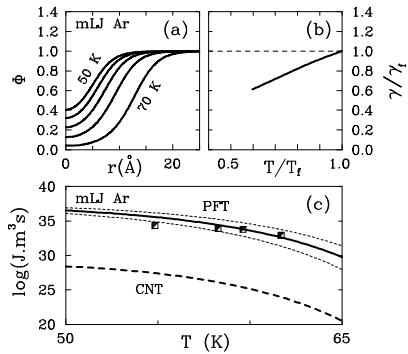}
}
\caption{
Nucleation in 
the modified Lennard-Jones system: (a) 
Radial phase field profiles for critical fluctuations at several
temperatures. (b) Relative interfacial 
free energy vs. reduced temperature. 
(c) Comparison of nucleation rates
predicted by the phase field theory (PFT), the classical 
sharp interface theory (CNT) and computer 
simulations (squares) \cite{bc95}. Short-dashed lines show 
the limits of nucleation rate allowed by the error of the 
interfacial free energy.  
$\epsilon = 137.07 k$ and $\sigma = 3.383$
 \AA $  $ are taken for Ar.
Below 61.7 K, the simulations increasingly
underestimate the true nucleation rate
due to an unknown equilibration period caused by quenching the
liquid to the nucleation temperature \cite{bc95}.
}
\label{mlj}
\end{figure}

To study the soft-impingement problem we introduce a non-conservative 
orientation field $\theta$, which is random in the liquid, and has 
a constant value between 0 and 1 in the crystal that determines 
crystal orientation in the laboratory frame. By this, we capture 
the feature that the short-range order in the solid and liquid are usually 
similar, with the obvious 
difference that the building 
units have a uniform orientation in the crystal, while their 
orientation fluctuates in the liquid. Following \cite{theta}, we assume 
that the grain boundary energy acts in the solid and is 
proportional to $| \nabla \theta |$. We realize this by 
adding $f_{ori} = {\cal M} | \nabla \theta |$ to $f_S$, where 
coefficient ${\cal M}$ is assumed to be independent of $c$. 
The respective 
equation of motion has the form 
$\partial \theta /\partial t = - M_{\theta} (\delta F/\delta \theta) 
+ \zeta_{\theta}$, yielding 
\begin{eqnarray}
{\frac{\partial \theta}{\partial \tilde t}} = {\frac{\xi M_{\theta} {\cal M}} 
{D_L}} \tilde \nabla \left\{ [1 - P(\phi)] {\frac{\tilde \nabla 
\theta}{| \tilde \nabla \theta |}} \right\} + \zeta_{\theta},
\end{eqnarray}
\noindent where $\zeta_{\theta} = \zeta_{\theta,0} P(\phi)$, and 
$M_{\theta} = M_{\theta,S} + P(\phi) (M_{\theta,L} - M_{\theta,S})$,
while subscripts S and L indicate the values for the bulk liquid 
and solid phases. When $\phi < 1$, Eq. (2) 
switches in orientational ordering, and chooses the value of $\theta$ 
that survives as the orientation of the particle, which 
then serves as the direction relative to which the anisotropy 
of $\gamma$  ($= \gamma_0 \left\{1 + s_0 cos[m (\vartheta - \theta)] \right\}$ 
for $m$-fold symmetry) is measured.
A similar model has been successfully applied for describing 
grain boundary dynamics \cite{theta,warr}. Unlike 
previous work, in our approach the orientation field $\theta$ is 
coupled to the phase field, and extends to the 
liquid phase where crystallographic orientation develops from
orientational fluctuations. While our model 
incorporates grain boundary dynamics, 
our primary interest is solidification, and $M_{\theta,S}$ is set 
so that grain rotation is negligible on the time scale of 
solidification.

Nucleation is incorporated into the simulations as follows. Method I.: 
By including white noise into the governing equations of amplitude 
that forces nucleation in the spatial and time windows used. Method II.: 
The simulation area is divided into domains according to the local 
composition. The time-independent solution is found for these 
compositions. Critical fluctuations of statistically correct numbers 
following Poisson distribution are placed into these areas in every 
time step. The added small-amplitude noise makes these critical 
fluctuations either grow or shrink. 

In nucleation-growth processes the transformed fraction often follows 
the Kolmogorov scaling, $X(t) = 1-exp\left\{-(t/t_0)^p\right\}$, 
where the "Kolmogorov exponent" $p$ is representative of the mechanism 
of the phase transformation, and is evaluated from the slope of the 
plot $ln[-ln(1-X)]$ vs. $ln t$. In this work $\phi < 0.5$ is used to 
define the transformed fraction.

\begin{figure}
\resizebox{60mm}{!}{
\includegraphics{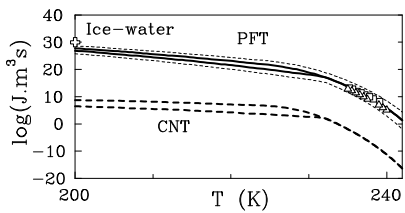}
}
\caption{Nucleation rate vs. temperature in the ice-water system.
The experimental results (symbols) are from \cite{icexp}. 
The branches below 232 K
indicate results obtained with physically reasonable upper and lower
estimates for the Gibbs free energy of undercooled water \cite{diff1}.
Notation as in Fig. \ref{mlj}.
}
\label{ice}
\end{figure}

First we apply the phase-field theory to predict nucleation rate in 3D.
In the {\it one-component} Lennard--Jones system, the nucleation 
rate \cite{bc95} and 
all the relevant physical properties are known 
from molecular dynamics simulations \cite{param1}. 
The radial phase field profiles (Fig. \ref{mlj}a) indicate that the critical
fluctuations are diffuse, and do not show bulk crystal properties for
undercoolings larger than 14 K. The predicted interfacial free energy 
(Fig. \ref{mlj}b) increases with temperature. While the phase-field 
predictions agree with results from computer simulations, those from the 
classical sharp interface theory differ from the experiments by 
eight to ten orders of magnitude. Similar results were obtained for the
ice-water system (Fig. \ref{ice}) with input data from \cite{param3}. 
Without adjustable parameters, a quantitative agreement has been achieved 
with computer simulations \cite{bc95} and experiment \cite{icexp},
proving the power of the phase-field technique
in attacking nucleation problems.

\begin{figure}
\resizebox{55mm}{!}{
\includegraphics{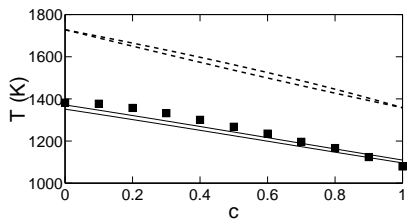}
}
\caption{Nucleation 
temperature vs. composition that the phase field theory predicts for the 
nearly ideal Cu--Ni system. Upper and lower solid lines correspond to 
nucleation rates of $10^{-4}$ and 1 drop$^{-1}$s$^{-1}$ for droplets 
of 6 mm diameter. 
The experimental data (squares) refer to electromagnetically levitated 
droplets \cite{whf88}. The calculated liquidus and solidus lines (dashed) are 
also shown.}
\label{bin}
\end{figure}

In the case of {\it binary alloys}, such a rigorous test cannot be 
performed since the input information available for the crystal-liquid 
interface is far less reliable. 
In the nearly ideal Cu--Ni system, the critical undercoolings 
computed for the realistic range of nucleation rates ($J = 10^{-4}$ 
to 1 drop$^{-1}$s$^{-1}$ for electromagnetically 
levitated droplets of 6 mm diameter) fall close to the experimental 
ones \cite{whf88} (Fig. \ref{bin}), indicating homogeneous nucleation. This 
contradicts the heterogeneous mechanism suggested earlier \cite{whf88} on 
the basis of Spaepen's value $\alpha_{HS} = 0.86$ \cite{sm78} for the 
dimensionless interfacial free energy for the hard-sphere system. 
(For definition of $\alpha$ see \cite{param1}.)
Recent computer simulations \cite{dl00}
yielded considerably smaller values $\alpha_{HS} 
= 0.51$ and $\alpha_{Ni} = 0.58$ 
($\approx 0.6$ we used), invalidating the earlier conclusion. 
These findings raise the possibility that homogeneous nucleation 
is more common in alloys than previously thought.

We turn now to the problem of {\it soft-impingement} that 
we investigate in 2D using the 
properties of Cu--Ni alloys. Owing to the known difficulties 
of phase field simulations 
due to different time and length scales of the fields, we used 
an enhanced interface thickness ($\delta = 41.6$ nm), 
a reduced interfacial free 
energy ($\alpha = 0.1$), and an increased diffusion coefficient
($\sim 100 \times D_L$). 
To ensure reasonable statistics and negligible influence of 
the periodic boundary conditions, the governing equations were 
solved numerically on a $7000 \times 7000$ grid under conditions
\cite{param2} that ensure interfaces consisting of more 
than ten grid points needed for numerical accuracy.

\begin{figure}
\resizebox{80mm}{!}{
\includegraphics{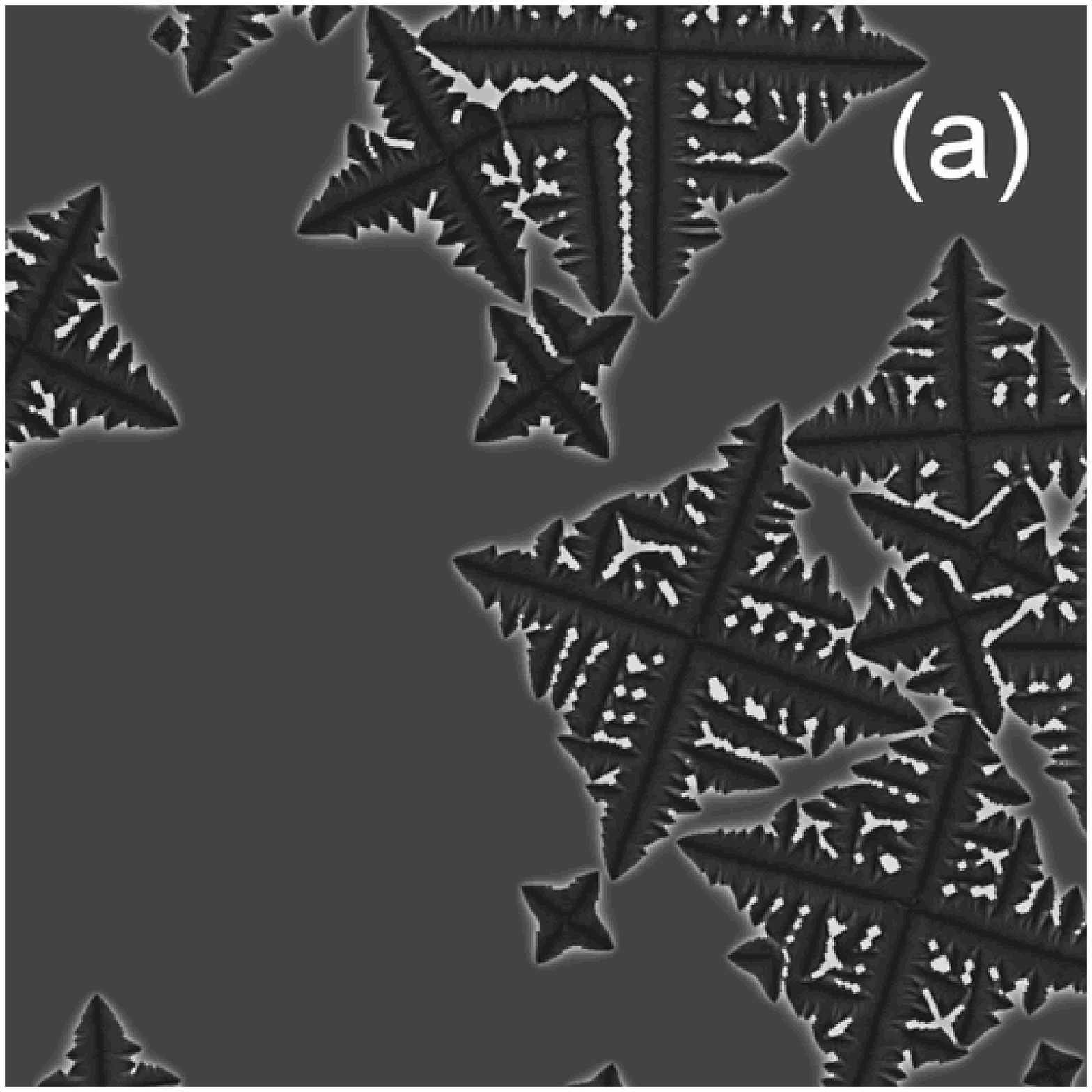}
\includegraphics{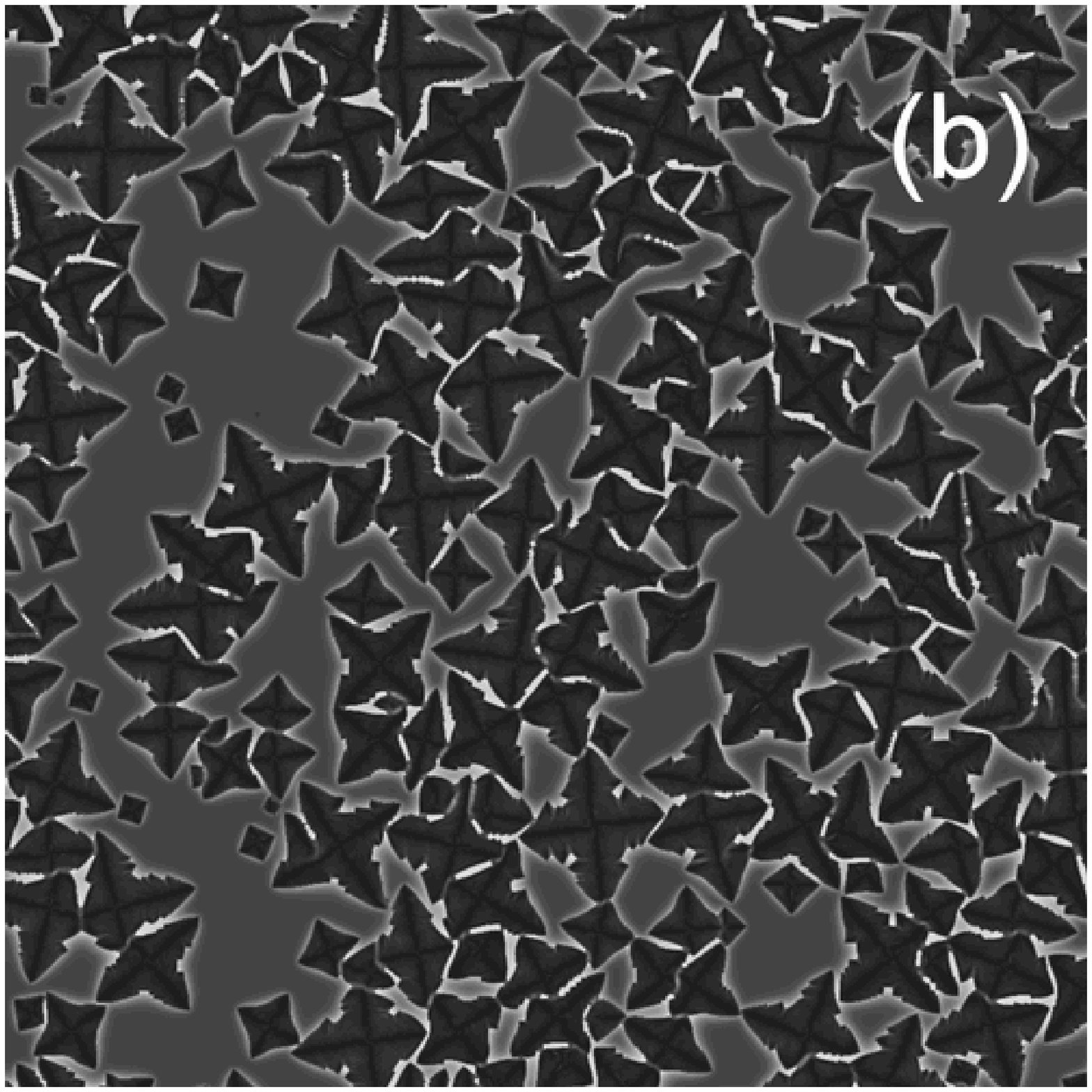}
\includegraphics{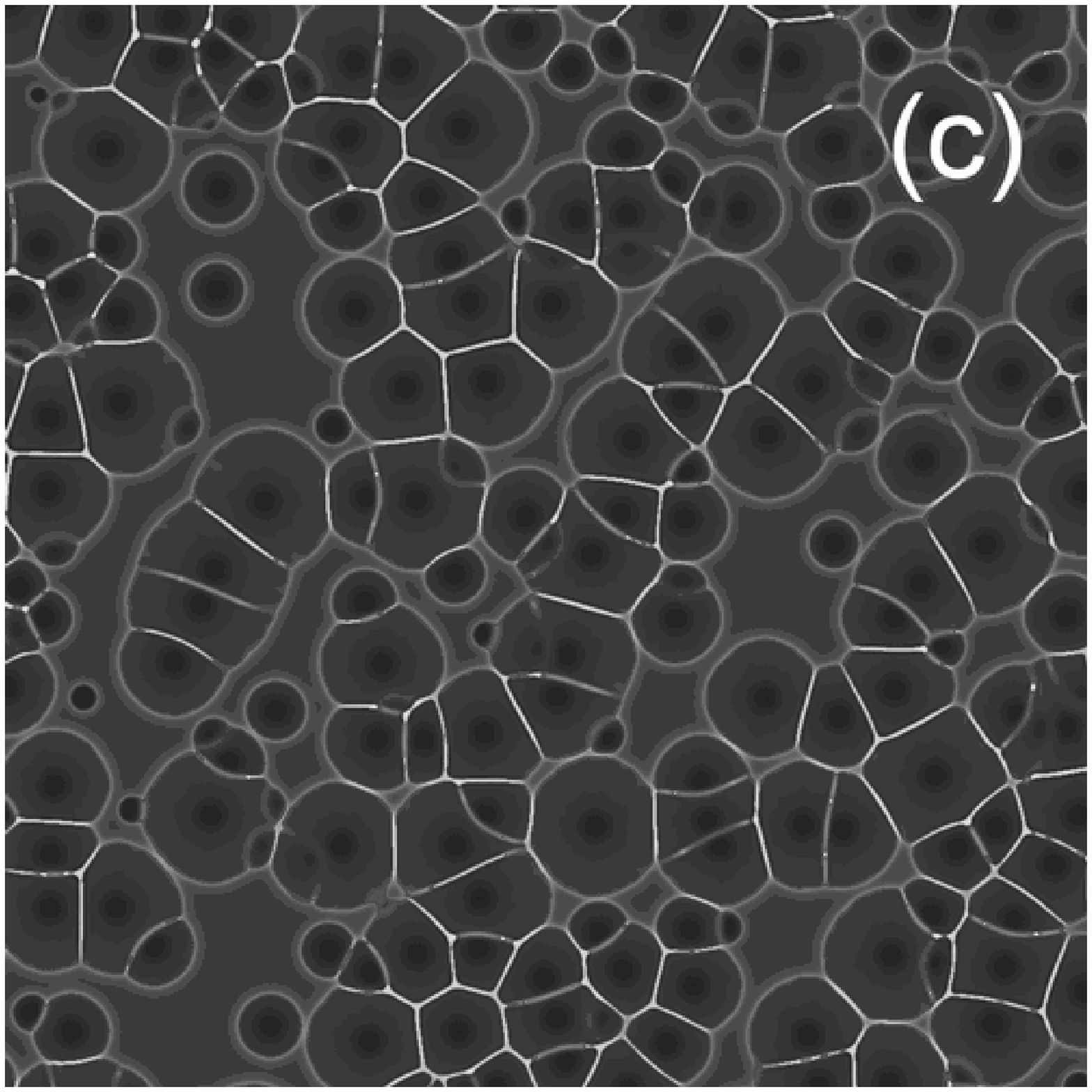}
}
\resizebox{26.7mm}{!}{
\includegraphics{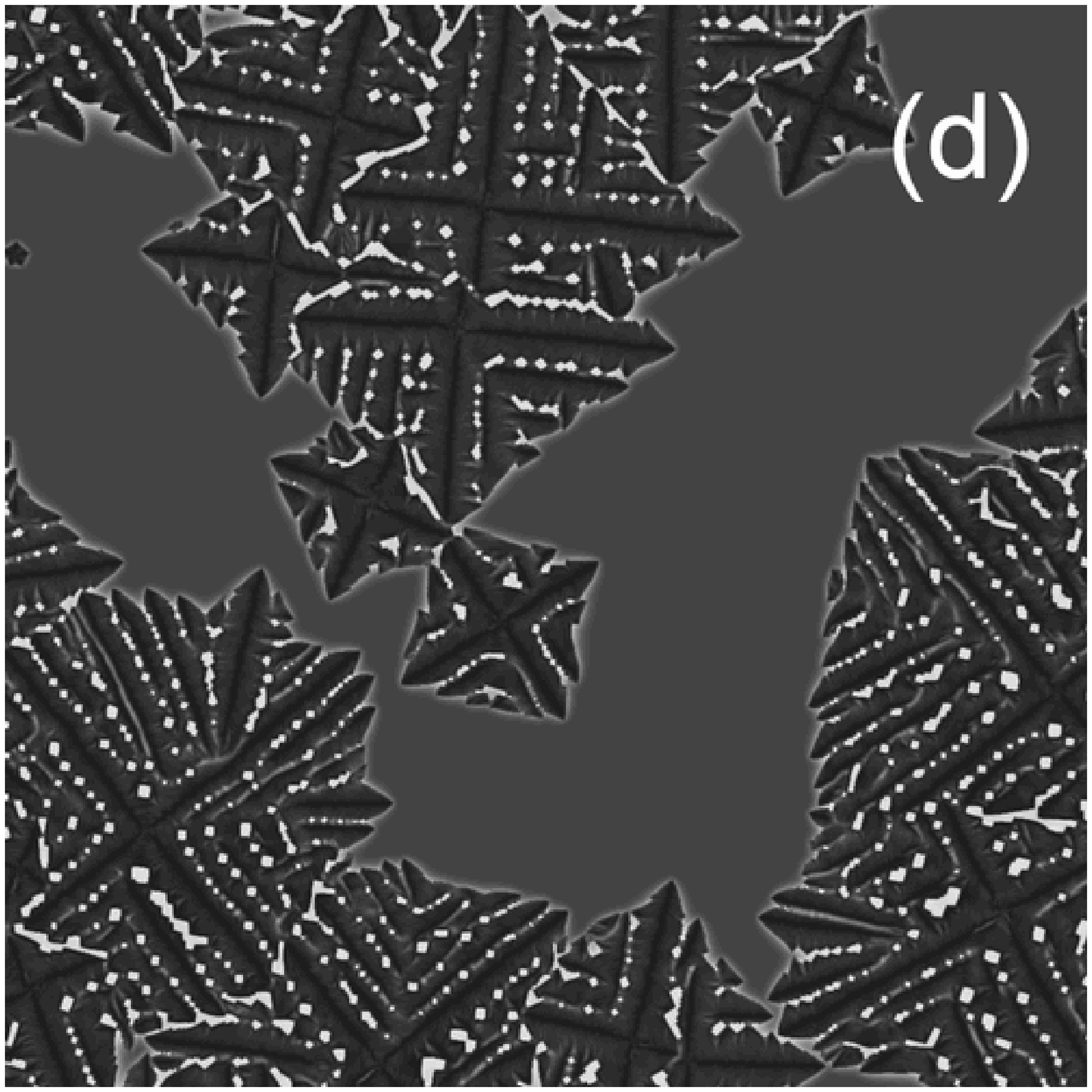}
}
\resizebox{52mm}{!}{
\includegraphics{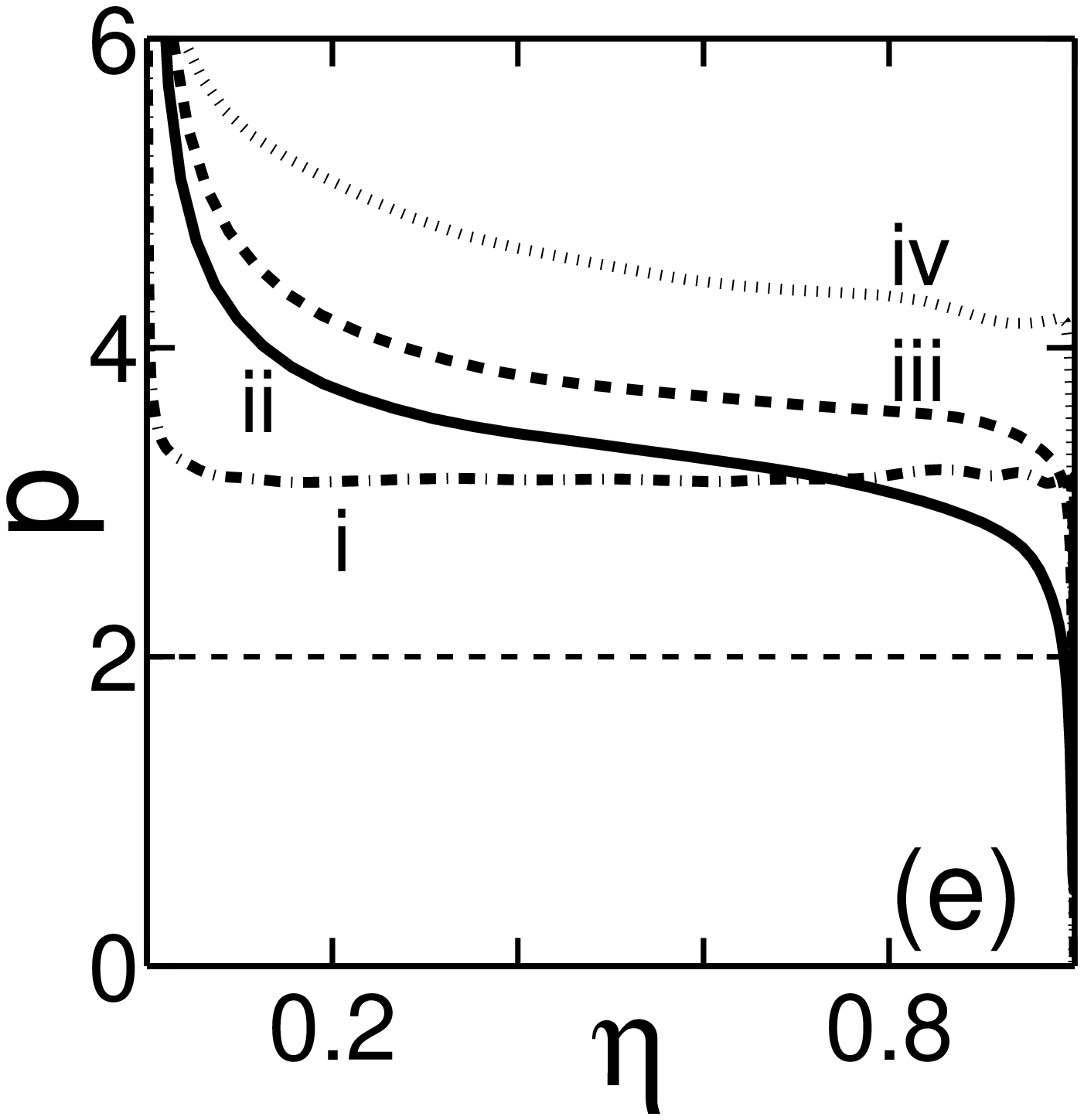}
\includegraphics{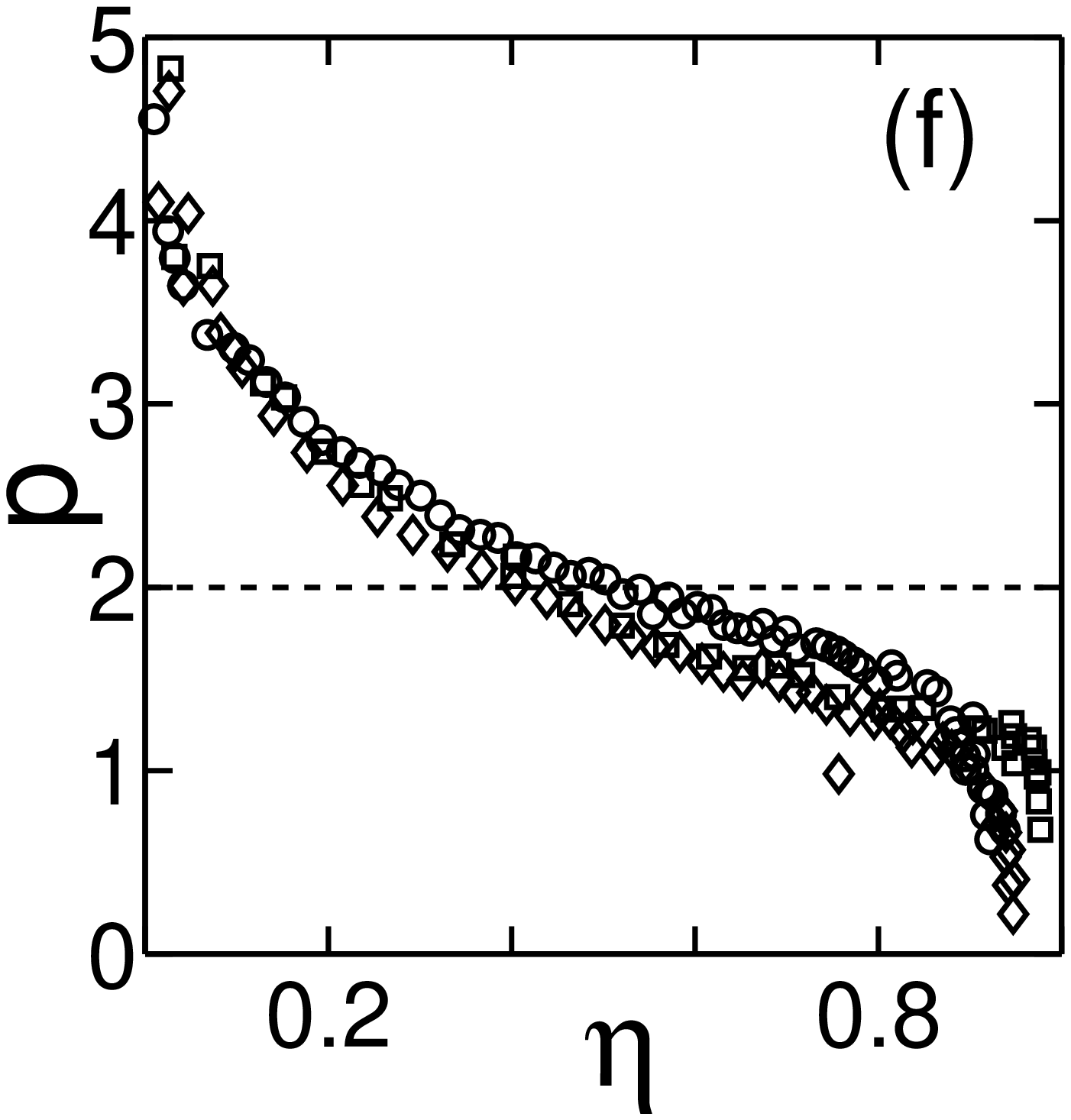}
}
\caption{Soft-impingement in the phase-field theory. (a)--(d) 
$1000 \times 1000$ segments of $7000 \times 7000$ snapshots for 
the concentration field in cases (i)-(iv), respectively, taken at $\tilde t/
\Delta \tilde t = 4000, 1500,$ 1500, and 6000 (black and white 
correspond to the solidus and liquidus); (e) 
Kolmogorov exponent vs. crystalline fraction, 
$\eta = X/X_{max}$, where $X_{max}$ is the final 
value of the crystalline fraction; (f) experimental results for the
crystallization of amorphous Fe$_{73.5}$Si$_{17.5}$CuNb$_3$B$_5$ \cite{nano}.}
\label{sim}
\end{figure}

We begin by comparing three simulations where the particles were 
nucleated by Method II.: (i) large anisotropy $s_0 = 0.25$ and small 
dimensionless nucleation rate $\tilde J = 0.49$ yielding 886 dendritic 
particles (the 
first large scale simulation of multiparticle dendritic
solidification) [Fig. \ref{sim}(a)]; (ii) $s_0 = 0.25$ with large nucleation 
rate $\tilde J = 24.5$ (10623 particles) [Fig. \ref{sim}(b)]; 
and (iii) isotropic growth with $\tilde J = 24.5$ (10528 particles) 
[Fig. \ref{sim}(c)]. The respective Kolmogorov exponents differ from 
$p = 2$ that standard references \cite{christ} assign for steady-state 
nucleation and diffusion controlled growth [Fig. \ref{sim}(d)]. 
Our prediction for dendritic solidification, $p \approx 3$,
obeys the relationship
$p = 1 + d$ (constant nucleation and growth rates in 
$d$-dimensions) confirmed 
experimentally \cite{dend}, that follows 
from the steady-state traveling tip solution of the diffusion equation. 
In cases (ii) and (iii), $p$ deviates from 2, as the formation of the 
diffusion layer is preceded by a transient period in which 
phase field mobility controls growth. This period appears as an effective 
delay of the diffusion controlled process, yielding an initially 
enhanced $p$ that decreases with increasing transformed fraction 
\cite{nano}. This effect is pronounced at large nucleation rates 
for which the delay is comparable to the solidification time. Indeed, 
qualitatively similar behavior is observed at the extreme nucleation 
rates that occur during the formation of nanocrystalline alloys \cite{nano}.

The introduction of large amplitude noise into the governing equations
(Method I.) leads to comparable results. For $s_0=0.25$ with 
$\zeta_{\phi,0}=0.015$, that yield $\sim 830$ dendritic particles 
[case (iv)], more irregular shapes are produced [Fig. 4(d)], 
and transient nucleation of ``induction time'' 
$\tau /\Delta \tilde t \approx 1400$ is observed. The latter leads to 
further increased $p$, that reduces to $\sim 3$ when replacing $\tilde t$
by $\tilde t - \tau$.
Due to numerical stability problems appearing at large noise amplitudes, 
Method I. can be applied far from equilibrium, where 
nucleation occurs in reasonable simulation time and area.

Summarizing, we demonstrated that the present phase field model is 
able to {\it quantitatively} describe crystal nucleation in one-component 
3D systems. The other predictions, including binary nucleation in 3D and 
transformation kinetics in 2D, are also consistent with experiment.

This work has been supported by the ESA Prodex Contract No. 14613/00/NL/SFe, 
by the Hungarian Academy of Sciences under contract Nos. OTKA-T-025139 and T-037323, by the EU grant HPMF-CT-1999-00132, and the ESA MAP Project No. AO-99-101.


\begin{thebibliography}{99}
\bibitem{param} 
$\epsilon_{i}^2= 6 \cdot 2^{1/2} \gamma_{i} \delta_{i} / T_{i}$ and 
$W_{i}= 12 \gamma_{i} / (2^{1/2} \delta_{i} T_{i})$, where 
$i =$ A or B. To satisfy the four relationships, we assume 
that $\epsilon^2$ is linearly temperature dependent.
\bibitem{param2}
The time and spatial steps were $\Delta \tilde t = 4.75 
\times 10^{-6}$ and 
$\Delta \tilde x = 6.25 \times 10^{-3}$, $\xi=2.1 
\times 10^{-4}$ cm and $D_L = 10^{-5}$ cm$^2$/s. Dimensionless mobilities 
$\xi^2 M_{\phi}/D_L = 0.9$, $\xi M_{\theta,L} {\cal M}/D_L$ = 720, and
$\xi M_{\theta,S} {\cal M}/D_L = 7.2 \times 10^{-4}$ were applied, 
while $D_S = 0$ 
was taken in the solid. The simulations were performed at 1574 K for 
$x = (C_{\infty} - C_s)/(C_l - C_s) = 0.2$, where $C_s = 0.399112$ 
and $C_l = 0.466219$ are the solidus and liquidus compositions.
White noises of amplitudes $\zeta_{\phi,0}=0.0025$, $\zeta_{\nabla c,0} = 0$,
and $\zeta_{\theta,0} = 0.25$ were used.
\end{thebibliography}
\end{document}